\documentclass[aps,showpacs,preprintnumbers,amsmath, amssymb]{revtex4}

\usepackage{float}
\usepackage{graphics,epsfig}
\usepackage{graphicx}
\usepackage{dcolumn}
\usepackage{bm}

\begin{document}
\baselineskip=0.8 cm

\title{{\bf Lower bound on the orbital period of Kerr-Newman black holes}}
\author{Yan Peng$^{1}$\footnote{yanpengphy@163.com}}
\affiliation{\\$^{1}$ School of Mathematical Sciences, Qufu Normal University, Qufu, Shandong 273165, China}

\vspace*{0.2cm}
\begin{abstract}
\baselineskip=0.6 cm
\begin{center}
{\bf Abstract}
\end{center}

Based on the orbital period of Kerr black holes, Hod proposed a conjecture that a general lower
bound on the orbital period may exist.
In this work, we examined this bound by exploring the orbital period of Kerr-Newman black holes using analytical
and numerical methods. By choosing different charge and spin of Kerr-Newman black holes,
we found a lower bound for the orbital period of Kerr-Newman black holes as $T(r)\geqslant 4\pi M$,
where $r$ is the orbital radius, $T(r)$ is the orbital period observed from infinity and $M$ is the black hole mass.
This bound is just the same as Hod's conjectured lower bound. So our results further demonstrated that Hod's lower bound
may be a general property in black hole spacetimes.

\end{abstract}

\pacs{11.25.Tq, 04.70.Bw, 74.20.-z}\maketitle
\newpage
\vspace*{0.2cm}

\section{Introduction}

The study of black hole spacetimes and the motion of particles within these spacetimes has been
a central topic in general relativity for a long time \cite{NG1,NG2,NG3,NG4,NG5}. The motion of test particles in black
hole spacetimes provides valuable insights into the structure of the spacetime geometry and has
implications for astronomical observations as well as gravitational wave studies \cite{PR1}-\cite{PR14}.
The continued exploration of black hole spacetimes is thus essential for
understanding of the extremely curved spacetimes.

An interesting and physical question is to search for the circular trajectory
with the shortest orbital period around black holes.
Hod analytically studied the minimum orbital period
measured by asymptotically flat space observers in backgrounds
of Schwarzschild black holes and Kerr black holes \cite{Hod1,Hod2}.
For non-rotating Schwarzschild black holes, it was found that the orbital
period satisfies  $T(r)\geqslant 6\sqrt{3}\pi M $ \cite{Hod1,Hod2}. For rotating
Kerr black holes, the situation is more complex due
to the presence of the black hole's angular momentum.
For Kerr black holes, Hod provided a rigorous analytical proof that
the orbital period $ T(r)$ cannot be less than the
lower bound \( 4\pi M \), where the bound can be approached for the maximally rotating Kerr black holes \cite{Hod1}.
Hod further conjectured that the lower bound \( 4\pi M \) on orbital periods may be a general property
in the spacetimes with compact objects in the center.
This idea is intriguing as it serves as a rule that governs how things move around compact objects.
In fact, this bound can be violated as Hod recently found that the orbital period can approach $2 \pi M$ around a super-spinning naked singularity \cite{Hod2} .
However, Hod's bound may be a general property in pure black hole spacetimes.
It needs further examination, especially in more complex black hole scenarios.
For Kerr-Newman black holes, the lower bound of orbital period may be also
related to the black hole's charge.
So it is interesting to study the lower bound of orbital period
in the background of charged rotating Kerr-Newman black holes.
In particular, it is meaningful to examine Hod's conjured lower bound
for more general Kerr-Newman black holes.

In this work, we firstly introduce the Kerr-Newman black
hole spacetimes. Then, we analysis the motion in the equatorial plane
and obtain the minimum orbital period.
With analytical and numerical methods, we prove that
the orbital period cannot be less than the lower bound \( 4\pi M \)
for different values of the black hole's mass $M$, charge $Q$ and angular momentum $a$,
which is in accordance with Hod's bound.
Finally, we summarize our main results in the last section.

\section{Calculating the lower bound on orbital period of Kerr-Newman Black Holes}

The Kerr-Newman black hole is a rotating and charged black hole in general relativity.
This type of black hole is characterized by black hole mass $M$, charge $Q$ and
angular momentum per unit mass $a$.
The geometry of the Kerr-Newman black hole is usually described with
the Boyer-Lindquist coordinates in the form \cite{NG2}:
\begin{equation}
ds^2 = -\left(1 - \frac{2Mr - Q^2}{\rho^2}\right) dt^2 - \frac{2a(2Mr - Q^2) \sin^2 \theta}{\rho^2} dt d\phi + \frac{\rho^2}{\Delta} dr^2 + \rho^2 d\theta^2
+\left[ r^2 + a^2 + \frac{a^2 (2Mr - Q^2) \sin^2\theta}{\rho^2} \right] \sin^2\theta  d\phi^2,
\end{equation}
where $\rho^2 = r^2 + a^2 \cos^2 \theta$, $\Delta = r^2 - 2Mr + a^2 + Q^2$.
The location of the black hole event horizon is determined by the condition $\Delta = r^2 - 2Mr + a^2 + Q^2=0$.
The outer and inner event horizon radius \( r_{\text{horizon}} \) of a Kerr-Newman black hole is given by the formula
$r_{\text{horizon}} = M \pm \sqrt{M^2 - a^2 - Q^2}$.

We obtain period of circular orbits in the equatorial plane following analysis in \cite{Hod1,Hod2,Yan Peng}.
On the equatorial plane, $\theta = \frac{\pi}{2}$, hence $\sin \theta = 1$
and $\cos \theta = 0$. It simplifies the expression of $\rho^2$ into $\rho^2 = r^2$.
For objects traveling in circular paths around such black holes,
the circular orbit condition is $dr = d\theta = 0$. Therefore, the metric can be transformed into:
\begin{equation}
ds^2 = -\left(1 - \frac{2Mr-Q^2}{r^2}\right) dt^2 - \frac{2a(2Mr - Q^2)}{r^2} dt d\phi + \left[r^2 + a^2 + \frac{a^2(2Mr - Q^2)}{r^2}\right] d\phi^2.
\end{equation}

As we are interested in the minimum orbital period, we focusing on objects approaching the speed of light.
The light speed condition signifies that the spacetime interval $ds^2$ is zero.
This is a fundamental requirement in general relativity for null geodesics,
which describe the paths that light follows in curved spacetime.
At the light speed limit, there is $ds^2 = 0$, and the relation (2) becomes:
\begin{equation}
-\left(1 - \frac{2Mr-Q^2}{r^2}\right) dt^2 - \frac{2a(2Mr - Q^2)}{r^2} dt d\phi + \left[r^2 + a^2 + \frac{a^2(2Mr - Q^2)}{r^2}\right] d\phi^2 = 0.
\end{equation}

Setting the coordinate $d\phi=2\pi$ to represent a full orbital period, and letting $dt=T(r)$
denote the time period, the original equation can be expressed as:
\begin{equation}
-\left(1 - \frac{2Mr-Q^2}{r^2}\right) T(r)^2 - \frac{4 \pi a(2Mr - Q^2)}{r^2} T(r) +4 \pi^2 \left[r^2 + a^2 + \frac{a^2(2Mr - Q^2)}{r^2}\right] = 0.
\end{equation}

Solving this equation, we obtain the orbital period:
\begin{equation}
T(r) = 2\pi \frac{\sqrt{r^2-2Mr+a^2+Q^2}-\frac{2Ma}{r}+\frac{aQ^2}{r^2}}{1-\frac{2M}{r}+\frac{Q^2}{r^2}}.
\end{equation}
This expression implies that the orbital period of
a circular orbit in the equatorial plane depends on the radius $r$,
black hole mass $M$, charge $Q$ and angular momentum $a$.

To find the minimum value of $T_{min}$, we calculate the derivative of the orbital
period with respect to the radial coordinate r. The derivative is given by:
\begin{equation}
\frac{dT}{dr} = \frac{2\pi r \left[ 2Q^4 + Q^2r(3r-7M ) + 2a^2(Q^2 - Mr) + r^2(6M^2 - 5Mr + r^2) - 2a(Q^2 - Mr)\sqrt{a^2 + Q^2 - 2Mr + r^2} \right]}{(Q^2 + r(r-2M ))^2 \sqrt{a^2 + Q^2 + r(r-2M)}}.
\end{equation}
We set \( \frac{dT}{dr} = 0 \) and solve it for the region outside the outer horizon.
This equation is too complex and numerical methods are needed.

We plot the normalized minimum orbital period $\frac{T_{min}}{M}$ as a function of the black hole mass $M$
for fixed values of the black hole spin parameter $a=0.5$ and charge $Q=0.5$ in Fig. 1.
As $M$ increases, the normalized minimum period $\frac{T_{min}}{M}$
also increases, indicating that larger black holes have longer normalized minimum orbital periods.
It can be seen from the picture that normalized minimum orbital periods is above $4 \pi\approx 12.56$.
For clarity, we also give the data in Table I.
The data shows that minimum orbital periods is indeed above $4 \pi\approx 12.56$.
In the case of $a=0.5$, $Q=0.5$ and $M=0.707109$, there is $M\thickapprox\sqrt{Q^2+a^2}+0.0000022$
nearly to an extreme black hole satisfying $M^2-Q^2-a^2=0$. For this case,
the normalized minimum orbital period is 13.3585, still above $4 \pi\approx 12.56$.

\begin{figure}
    \centering 
    \includegraphics[scale=0.8]{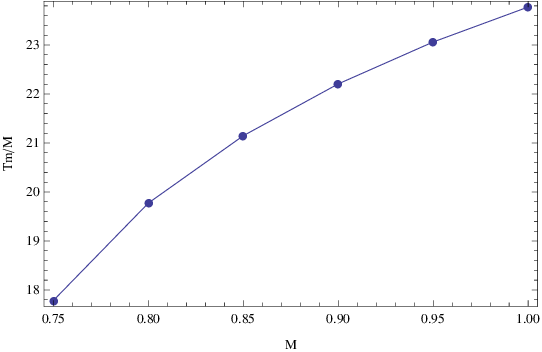} 
    \caption{$\frac{T_{min}}{M}$ as a function of $M$ with $a=0.5$ and $Q=0.5$.} 
\end{figure}

\begin{table}[htbp] 
    \centering 
     \caption{$\frac{T_{min}}{M}$ with $a=0.5$, $Q=0.5$ and various $M$.} 
     \renewcommand{\arraystretch}{2.5} 
    \begin{tabular}{|c|c|c|c|c|c|c|c|} 
        \hline 
        $M$ & 0.707109  & 0.75 & 0.80 & 0.85 & 0.90 & 0.95 & 1.00  \\ 
        \hline 
        $\frac{T_{min}}{M}$ & 13.3585 & 17.79 & 19.77 & 21.15 & 22.21 & 23.06 & 23.78  \\ 
        \hline 
    \end{tabular}
    \label{tab:my_label} 
\end{table}

We represent $\frac{T_{min}}{M}$ as a function of $a$ with $M=1.0$ and $Q=0.5$ in Fig. 2. For a fixed
black hole mass $M=1.0$ and charge $Q=0.5$, increasing the spin parameter $a$ leads to a decrease in the normalized
minimum orbital period $\frac{T_{min}}{M}$.
It can be seen from the picture that $\frac{T_{min}}{M}$ is always above the bound $4 \pi\approx 12.56$.
Table II provides the detailed data of minimum orbital period $\frac{T_{min}}{M}$
for different values of the spin parameter $a$, with the black hole mass $M$ and charge $Q$ kept as constants 1.0 and 0.5, respectively.
The data illustrates clearly that the minimum orbital period is larger than the bound $4 \pi\approx 12.56$.
In the case of $M=1.0$, $Q=0.5$ and $a=0.866015$, there is $a\thickapprox\sqrt{M^2-Q^2}-0.00001$
nearly to an extreme black hole and the normalized minimum orbital period is 12.7459 above the bound $4 \pi\approx 12.56$.

\begin{figure}
    \centering 
    \includegraphics[scale=0.8]{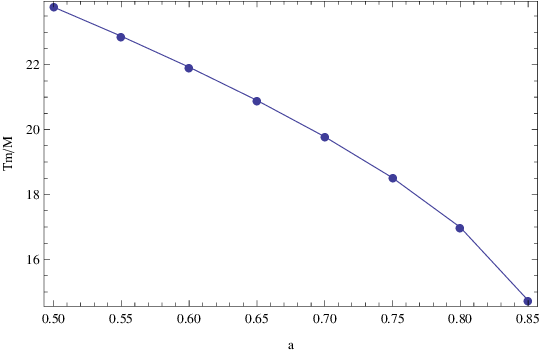} 
    \caption{$\frac{T_{min}}{M}$ as a function of $a$ with $M=1.0$ and $Q=0.5$.} 
\end{figure}

\begin{table}[htbp] 
    \centering 
     \caption{$\frac{T_{min}}{M}$ with M=0.5, Q=0.5 and various $a$.} 
     \renewcommand{\arraystretch}{2.5} 
    \begin{tabular}{|c|c|c|c|c|c|c|c|c|c|c|} 
        \hline 
     $a$ &  0.50 & 0.55 & 0.60 & 0.65 & 0.70 & 0.75& 0.80&0.85  & 0.866015 \\ 
        \hline 
       $\frac{T_{min}}{M}$ & 23.78 & 22.88 & 21.92 & 20.90 & 19.78 & 18.52&16.99&14.74  & 12.7459 \\ 
        \hline 
    \end{tabular}
    \label{tab:my_label} 
\end{table}

We study the effects of black hole charge $Q$ on the minimum orbital period $\frac{T_{min}}{M}$
normalized by the black hole mass $M$ in Fig. 3. For a fixed mass $M=1.0$ and spin parameter $a=0.5$,
increasing the charge $Q$ results in a decrease in the normalized minimum orbital period $\frac{T_{min}}{M}$.
This demonstrates that black holes with larger charge possess smaller normalized minimum orbital period.
Table III lists the normalized minimum orbital period $\frac{T_{min}}{M}$
for various values of the black hole's charge $Q$, with the mass $M=1.0$ and spin parameter $a=0.5$ respectively.
The data shows clearly that the normalized minimum orbital period is larger than the bound $4 \pi\approx 12.56$.
The particular case of $M=1.0$, $a=0.5$ and $Q=0.866015 \thickapprox \sqrt{M^2-a^2}-0.00001$
is almost an extreme black hole
and the normalized minimum orbital period is 15.7233 above $4 \pi\approx 12.56$.

\begin{figure}
    \centering 
    \includegraphics[scale=0.8]{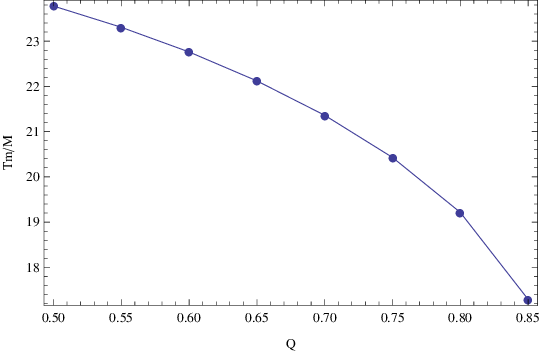} 
    \caption{$\frac{T_{min}}{M}$ as a function of $Q$ with $M=1.0$ and $a=0.5$.} 
\end{figure}

\begin{table}[htbp] 
    \centering 
     \caption{$\frac{T_{min}}{M}$ with $M=1$, $a= 0.5$ and various $Q$.} 
     \renewcommand{\arraystretch}{2.5} 
    \begin{tabular}{|c|c|c|c|c|c|c|c|c|c|} 
        \hline 
        $Q$  &  0.50 & 0.55 & 0.60 & 0.65 & 0.70 & 0.75& 0.80&0.85  & 0.866015 \\ 
        \hline 
      $\frac{T_{min}}{M}$ & 23.78 & 23.31 & 22.76 & 22.12 & 21.36 & 20.43&19.21&17.29  & 15.7626 \\ 
        \hline 
    \end{tabular}
    \label{tab:my_label} 
\end{table}

By choosing various parameters, we try to examine
Hod's conjectured bound in Kerr-Newman black holes.
The data of dimensionless ratio $\frac{T_{min}}{M}$
in Tables I, II and III and the curves in Fig.1, Fig.2 and Fig.3
all support the Hod's conjecture that the orbital period of
black holes possess a lower bound $4\pi M $.
In fact, Hod's original conjuncture was supposed to hold
for all compact objects and in contrast, Hod recently found that
the orbital period can approach $2 \pi M$ around a super-spinning naked singularity \cite{Hod1,Hod2}.
However, if we just consider black holes spacetimes, Hod's conjuncture still holds even
including black hole charge and angular momentum.
These results suggest that Hod's conjure may be a general property
for black hole spacetimes in the form:
\begin{equation}
T(r) \geqslant 4\pi M,
\end{equation}
where the bound can be approached for the maximally rotating Kerr black hole.

\section{Conclusions}

Based on earlier work on Kerr black holes, Hod conjectured that there might be a universal lower
bound $4\pi M$ on the orbital period, which sill needs further
examination in more complex black hole scenarios.
In this work, we examined Hod's conjecture in the
Kerr-Newman black hole background. Using a combination of analytical and numerical methods,
we found that the orbital period for these
black holes is indeed bounded by $T(r) \geqslant 4\pi M $. Here, $ r$ is the orbital radius,
$T(r)$ is the period observed from far away, and $M$ is the black hole's mass.
This result matches exactly with Hod's original conjectured bound. It suggests that
no matter the black hole spins fast or has a significant
charge, the shortest possible orbit can't get shorter than $4\pi M $. This finding provides more evidence that Hod's lower bound might be
a fundamental feature of all black hole spacetimes.

\begin{acknowledgments}

This work was supported by the Shandong Provincial Natural Science Foundation of China under Grant
No. ZR2022MA074. This work was also supported by a grant from Qufu Normal University
of China under Grant No. xkjjc201906.

\end{acknowledgments}

\end{document}